\documentclass[prl,aps,twocolumn,superscriptaddress]{revtex4}

\usepackage{graphicx,color}
\usepackage{dcolumn}
\usepackage{bm}
\usepackage{amssymb}

\begin{document}



\title{Spin superstructure and noncoplanar ordering in metallic pyrochlore magnets with degenerate orbitals}

\author{Gia-Wei Chern}
\affiliation{Department of Physics, University of Wisconsin, Madison, Wisconsin 53706, USA}

\author{Cristian D. Batista}
\affiliation{Theoretical Division, Los Alamos National Laboratory, Los Alamos, New Mexico 87545, USA}

\date{\today}

\begin{abstract}
We study double-exchange  models with itinerant $t_{2g}$ electrons in  
spinel and pyrochlore crystals. In both cases the localized spins  
form a network of corner-sharing tetrahedra. We show
that the strong directional dependence of $t_{2g}$ orbitals leads to unusual 
Fermi surfaces that induce spin superstructures and noncoplanar orderings
for a weak coupling between itinerant electrons and  localized spins. 
Implications of our results to ZnV$_2$O$_4$ and Cd$_2$Os$_2$O$_7$ are also discussed.
\end{abstract}

\maketitle

Orbital degrees of freedom have attracted much attention due to their crucial role 
in the stability of many unusual phases of correlated materials \cite{imada}.
In particular, the presence of degenerate orbitals in frustrated magnets
can lift the spin degeneracy through various spin-orbital interactions.
For Mott insulators with spins residing on a frustrated lattice, 
such as kagome or pyrochlore, geometrical constraints prevent spins from reaching a 
simple N\'eel order. The occurrence of long-range orbital order due to either Jahn-Teller 
distortion or orbital exchange reduces the magnetic frustration by creating disparities 
between nearest-neighbor (NN) exchange constants and paves the way for magnetic ordering.

However, some of the magnetic orders observed in geometrically frustrated compounds are difficult
to explain starting from the strongly coupled Mott-insulator regime. For example, several vanadium spinels 
\cite{lee,reehuis,bella} exhibit a complicated magnetic structure 
with $\uparrow\uparrow\downarrow\downarrow\cdots$ collinear ordering along certain chains that is  
very puzzling from the viewpoint of localized spin models. Below we shall provide a simple 
explanation for the observed spin superstructures based on a double-exchange (DE) model which 
takes into account orbital degeneracy. The DE model arises naturally for 
multiband compounds in which a narrow band of localized electrons coexists with a wider 
band of itinerant electrons. It can also be viewed as a mean-field approximation to the 
Hubbard Hamiltonian. A well studied case is the DE model with itinerant $e_g$ electrons  
on the cubic lattice \cite{brink}. This model has been shown to describe the rich physics
of  colossal magnetoresistance in perovskite manganites.

Recently there has been tremendous interest in DE models on frustrated lattices 
\cite{ikoma,ikeda,martin,akagi,kato,kumar,motome,chern}. The Fermi surface geometry plays
a crucial role \cite{martin,chern} in the nonlocal effective spin-spin interaction that
results from integrating out the itinerant electrons in the weak-coupling regime. 
The magnetic structures stabilized by itinerant electrons are thus often difficult 
to understand using short-range spin models. 
For example, an unusual noncoplanar magnetic order, in which spins on different sublattices point 
toward the corners of a tetrahedron, is shown to appear in different coupling regimes and various 
commensurate filling fractions on the triangular lattice \cite{martin,akagi,kato,kumar}.
Recent investigations of DE models on pyrochlore lattice also reveal interesting
behaviors such as electronic phase separation \cite{motome} and a complex noncoplanar order \cite{chern} 
at quarter-filling. However, most of these studies ignore the orbital dependence and consider only isotropic
electron hopping.

\begin{figure}
\includegraphics[width=0.9\columnwidth]{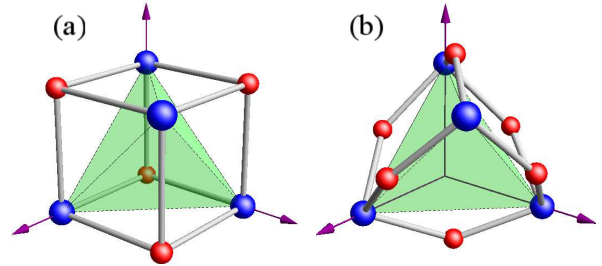}
\caption{\label{fig:lattice} A unit cell of the pyrochlore lattice and the 
configuration of local oxygen octahedra in (a) spinel and (b) pyrochlore crystals.
The blue and red balls denote the (B-site) transition-metal and oxygen ions, respectively.}
\end{figure}

In this Letter we examine DE models with itinerant $t_{2g}$ electrons
in both spinel and pyrochlore structures with general formulas  AB$_2$O$_4$ and A$_2$B$_2$O$_7$, respectively.
We consider localized classical spins residing on the B-sites of the crystal
which form a 3D network of corner-sharing tetrahedra.
The O$_6$ octahedron surrounding the B-sites creates a cubic crystal field which splits 
the $d$-levels into the $e_g$ doublet and the lower-energy $t_{2g}$ triplet. 
The strong dependence of electron hopping on orbital orientation leads to
peculiar Fermi surfaces in both cases. In particular, the electron subsystem reduces to 
a set of cross-linking Kondo chains in spinels. We show that this feature leads to a weak-coupling instability
towards the previously mentioned $\uparrow\uparrow\downarrow\downarrow$ superstructure in vanadium spinels.
Fermi surface nesting of different origin leads to noncoplanar all-in-all-out magnetic order in pyrochlores, which is a candidate state for the 
intermediate insulating phase of Cd$_2$Os$_2$O$_7$.

{\it Frustrated Kondo-chains in spinels}. In the spinel structure, a common quantization axis
can be defined for $t_{2g}$ electrons at all crystal B-sites [Fig.~\ref{fig:lattice}(a)].
The shape of the $t_{2g}$ orbitals is such that the strongest overlap is between the same orbitals 
along a particular NN direction, e.g., between two $d_{xy}$ orbitals 
along either a $[110]$ or $[1\bar 1 0]$ bonds in the $xy$ plane. Keeping only this dominant term,
electrons in a given orbital state can only hop along the corresponding $\langle 110 \rangle$ chain
on the pyrochlore lattice. We thus divide these chains into three types, $yz$, $zx$, 
and $xy$, depending on the active orbitals along the chain. Since the kinetic energy  preserves
the orbital flavor, the Hamiltonian is a sum of contributions from different 
orbital sectors:  ${\bar H} = \sum_{m} H_{m}$, where
\begin{eqnarray}
	\label{eq:H0}
	H_{m} = -t \sum_{\langle ij \rangle \parallel m}
	\left(c^{\dagger}_{i m \alpha} c^{\phantom{\dagger}}_{j m \alpha} + \mbox{h.c.}\right) 
      - {\tilde J}_H \sum_{i} \mathbf S_i\cdot {\mathbf s_{i m}},
\end{eqnarray}
$m=xy, yz, zx$ and ${\mathbf s_{i m}} = \frac{1}{2} \sum_{\alpha \beta} 
c^{\dagger}_{i m \alpha}\bm\sigma^{\phantom{\dagger}}_{\alpha\beta} c^{\phantom{\dagger}}_{i m \beta}$. 
Here the first term describes NN hopping of $t_{2g}$ electrons along a $\langle 110 \rangle$ 
chain of type $m$; $t$ is the dominating $dd\sigma$ transfer integral. $c^{\dagger}_{i m \alpha}$ is 
the creation operator for $d$-electrons at site $i$ with orbital flavor $m = xy, yz, zx$ 
and spin $\alpha = \uparrow,\, \downarrow$.  The second term in Eq.~(\ref{eq:H0})
describes an effective on-site Hund's coupling between $t_{2g}$ electrons and 
localized classical spins $\mathbf S_i$ (with normalization $|\mathbf S_i|$=1). Regarding model~(\ref{eq:H0})
as a mean-field approximation for a three-band Hubbard Hamiltonian that has the same kinetic 
energy term as ${\bar H}$, the effective coupling constant 
is ${\tilde J}_H = 4(U/9+4J_H/9) |\langle {\mathbf s_i} \rangle|$
where $U+J_{H}$ is the Coulomb repulsion between two electrons in the same orbital and $J_H$ is the bare Hund's coupling \cite{note1}.

$\bar H$ models a collection of  ferromagnetic (FM) 
Kondo chains coupled together by the local moments. While a classical Kondo chain is a relatively simple 
system, the fact that each spin is shared by three chains with different orbitals introduces geometric 
frustration. Numerical methods such as Monte Carlo (MC)  become very inefficient for conventional 
 3D DE models because the dimension of the electron Hamiltonian  
to be diagonalized for each spin update scales as $L^3\times L^3$ for systems with linear size $L$. 
On the contrary, for $\bar H$, one only needs to diagonalize matrices 
whose dimension scales as $L\times L$ for the three Kondo chains intersecting at the updated spin. 
The reduced dimensionality of the problem thus allows for studying the ground states 
of $\bar H$ with the aid of large-scale MC simulations.

\begin{figure}[t]
\includegraphics[width=0.98\columnwidth]{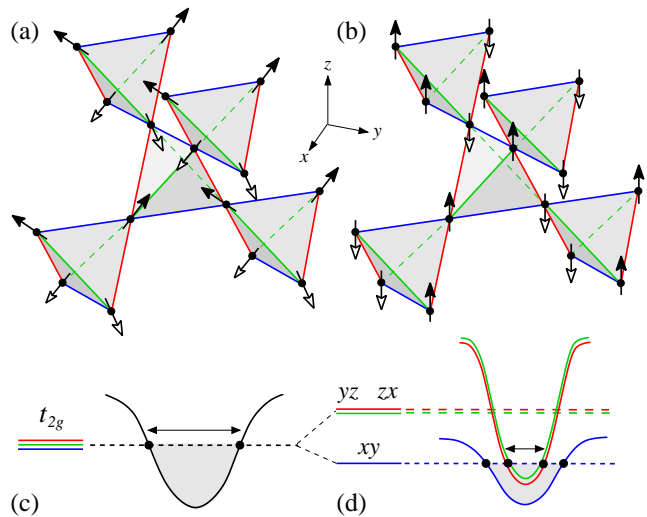}
\caption{\label{fig:m-order1} (a) The noncoplanar all-in all-out magnetic order and (b)
collinear spin superstructure with wavevector $\mathbf Q = (0,0,2\pi/a)$ in spinels.
Note the $\uparrow\uparrow\downarrow\downarrow\cdots$
order along  $yz$ and $xz$ chains and a $\uparrow\downarrow\uparrow\downarrow\cdots$ N\'eel order on 
$xy$ chains in~(b). 
Panels (c) and (d) show the corresponding electron band structures of the three 1D chains 
with different orbitals. }
\end{figure}

We first consider the case with three $d$-electrons per site. The electron energy
is minimized by placing one electron at each of the three different 1D bands, giving rise to
half-filled Kondo chains with a Fermi wavevector $k_F = \pi/2l$ [Fig.~\ref{fig:m-order1}(c)],
where $l$ is the NN distance.
The two Fermi points are nested by a commensurate wavevector $q_{1/2} = 2k_F = \pi/l$, 
leading to magnetic  N\'eel order in the presence of Hund's coupling. 
However, direct inspection shows that such a collinear N\'eel order cannot be simultaneously 
attained on all chains of the pyrochlore lattice. Instead, MC simulations
on $L=8$ lattices  (with $16L^3$ spins)  show that the total energy is minimized
by the noncoplanar all-in-all-out spin order shown in Fig.~\ref{fig:m-order1}(a).
The magnetic order of each chain consists of FM 
and  staggered components, which are perpendicular to each other.
It is worth noting that any global rotation of this noncoplanar spin ordering leads to 
another ground state due to the SU(2) invariance of ${\bar H}$.

The situation is more complicated for transition metals with two $d$-electrons per site
like the vanadium spinels AV$_2$O$_4$ where A $= $ Zn, Cd, or Mg. In ideal
cubic spinel, equal distribution of electrons among the Kondo chains corresponds 
to 1/3 filling fraction. The classical ground state of a single Kondo chain at $1/3$-filling
has a $\uparrow\uparrow\downarrow$ magnetic order with a period of $3\,l$. Again such a
simple arrangement of spins is precluded by geometric frustration. Our numerical
minimization on large finite systems yields a 3D noncoplanar magnetic
order with wavevector $\mathbf Q = \frac{2\pi}{a} \left ( \frac{1}{3},\frac{1}{3},1 \right)$,
where $a = 2\sqrt{2} \,l$ is the length of a conventional cubic unit cell;
the extended magnetic unit cell contains 108 spins.

{\it Collinear superstructure in vanadium spinels}.
Instead of the above complex order which preserves orbital degeneracy, experiments showed that 
vanadium spinels undergo a cubic-to-tetragonal structural transition with lattice 
constants $c < a = b$ \cite{lee,reehuis,bella}. 
Contrary to the elongation, which is favored by a Jahn-Teller ion with two $t_{2g}$ electrons,
the observed tetragonal compression can be understood as originating from the band Jahn-Teller 
instability. The lattice distortion results in a crystal-field splitting of the $t_{2g}$ levels 
as shown schematically in Fig.~\ref{fig:m-order1}(d). 
With two $d$-electrons per site, the lower energy $xy$ orbital is always occupied by one electron. 
Since the lattice distortion also reduces the hopping integral of $xy$ electrons, we assume that
the $d$-electrons in the $xy$ orbital are localized and constitute the local moments $\{\mathbf S_i\}$.
Therefore, we consider the DE Hamiltonian:
\begin{eqnarray}
	\label{eq:H_DE}
	H_{DE} =  \sum_{m = yz, zx} H_m 
	+ J_{AF} \sum_{\langle ij \rangle \parallel xy} \mathbf S_i\cdot\mathbf S_j.
\end{eqnarray}
Here $J_{AF} = 4 t^2_{xy}/U$ is the exchange constant between localized spins along
the $[110]$ and $[1\bar 1 0]$ chains. As the magnet is cooled, 
antiferromagnetic (AFM) spin correlations develop first along these chains as indeed observed \cite{lee}. 
A long-range 3D magnetic order resulting from interactions between different spin chains sets in 
at a lower temperature \cite{lee,reehuis,bella}. However, the crossing-chain coupling is 
geometrically frustrated if only  NN spin interactions are  taken into account \cite{tsunetsugu,tcher}.

Here we provide a simple picture of the unusual magnetic order of these vanadates based
on the DE model~(\ref{eq:H_DE}). Because the other $d$ electron can occupy either $yz$ or $zx$
orbitals, the corresponding bands are both 1/4 filled. In presence of Hund's coupling, $J_H$, 
the usual Fermi-point nesting thus leads to the formation of $\uparrow\uparrow\downarrow\downarrow\cdots$ superstructure with $q_{1/4}=2k_F=  \pi/2 l$ on both $yz$ and $zx$ chains. 
In the weak-coupling regime, this collinear  ordering (same amplitude for $\pm q_{1/4}$) 
is always more stable than the single-$q$ spiral order  because 
both wavevectors, $q_{1/4}$ and $-q_{1/4}$, are required to gap the two Fermi points of each chain.
The corresponding 3D collinear magnetic order [Fig.~\ref{fig:m-order1}(b)] 
characterized by wavevector $\mathbf Q = (0,0,1)$ is consistent with the experiments
(we shall from now on express the wavevectors in unit of $2\pi/a$ for convenience).
The mechanism for the formation of this magnetic order is similar to the orbitally induced
Peierls instability in the spinel MgTi$_2$O$_4$ \cite{khomskii}.

Although the above collinear spin order can also be explained within a local spin picture,
an {\it ad hoc} third-neighbor AFM  exchange has to be introduced in order to stabilize 
the $\uparrow\uparrow\downarrow\downarrow$ structure along $yz$ and $zx$ chains \cite{tsunetsugu}. 
On the other hand, our approach based on the itinerant DE model provides a natural explanation 
for the formation of these superstructures in  absence of orbital order. 
In addition, recent {\it ab initio}  calculations and experimental studies indicated that 
some vanadium compounds are indeed close to the metal-insulator transition \cite{pardo,giovannetti}, 
giving further  support to the itinerant picture adopted here.
Although the above conclusion is valid only for weak $J_H/t$, and is based
on a mean-field treatment of the multi-band Hubbard model, a more exact calculation that takes
into account the electron correlations gives a consistent result which will be 
presented elsewhere.

{\it Noncoplanar magnetic order in metallic pyrochlore}. We now turn to DE
model with degenerate orbitals on the pyrochlore structure [Fig.~\ref{fig:lattice}(b)].
Our theory provides a plausible explanation for magnetic ordering and metal-insulator
transition in the pyrochlore oxide Cd$_2$Os$_2$O$_7$. This compound undergoes 
a continuous metal-insulator transition at $T_{MI} \approx 225$ K \cite{sleight,mandrus,koda}.
The resistivity increases by 3 orders of magnitude upon cooling below $T_{MI}$.
The transition is accompanied by a sharp reduction of magnetic susceptibility, indicating
the occurrence of AFM order \cite{mandrus}. The specific-heat anomaly at $T_{MI}$
is found to be well described by a mean-field BCS-type phase transition.
The electron activation energy obtained from resistivity measurements also exhibits 
a BCS-like behavior near $T_{MI}$ \cite{mandrus}.

These experimental observations justify a mean-field approach for the metal-insulator
and magnetic transition in Cd$_2$Os$_2$O$_7$. As discussed above, the mean-field approximation 
reduces the multiband Hubbard model to the following DE model:
\begin{eqnarray}
	\label{eq:H2}
	{H}_{MF} = -\sum_{ij}\sum_{mn,\alpha} t^{mn}_{ij}\, c^{\dagger}_{i m \alpha} 
	c^{\phantom{\dagger}}_{j n \alpha} 
	 - {\tilde J}_H \sum_i\sum_m \mathbf S_i\cdot {\mathbf s_{i m}}.\,\,
\end{eqnarray}
Here the orbital index $m$ refers to the quantization axes of the {\it local} crystal fields
which are different in the four nonequivalent crystal B-sites [Fig.~\ref{fig:lattice}(b)].
Contrary to the case of spinels, the orbital flavor is not conserved by the kinetic term. 
To obtain the hopping matrix, we expand the $t_{2g}$ 
orbital wavefunction at a given sublattice $s$ in the basis of  common coordinates 
for the cubic pyrochlore: $|\phi^{(s)}_{m}\rangle = a^{s}_{m k} |\phi_k\rangle$.
The details of the transformation coefficients $a^{s}_{m k}$ can be found
in Ref.~[\onlinecite{tomizawa}]. The resulting hopping matrix is 
$
	t_{ss'}^{mn} =  \sum_{kl} a^s_{m k}\, a^{s'}_{n l} \,
	\langle \phi_{k}|H_t |\phi_{l}\rangle.
$
Here the transfer integral $\langle \phi_{k}|H_t|\phi_{l}\rangle$ is expressed
using the Slater-Koster (SK) parameters \cite{SK}.

\begin{figure}
\includegraphics[width=0.98\columnwidth]{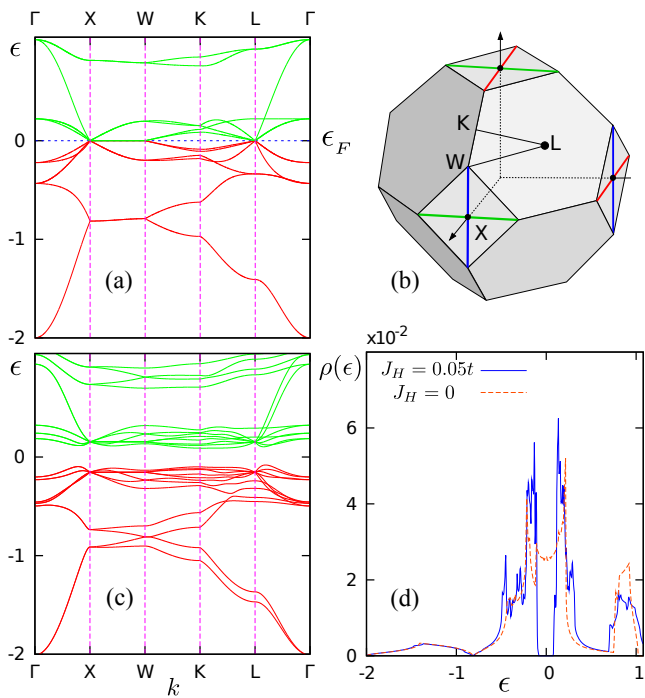}
\caption{\label{fig:ek} (a) Band structure of the orbital-dependent tight-binding
model in the pyrochlore crystal structure. At half-filling, the Fermi level is at $\epsilon = 0$.
(b) First Brillouin zone of the fcc lattice which is the underlying Bravais lattice
of the pyrochlore crystal. The Fermi `surface' for a half-filled band include
three sets of Fermi lines (the XW segment in (a)) and Fermi points
at the L points. (c)~Band structure in the presence of nonzero Hund's constant ${\tilde J}_H = 0.05t$.
(d) Calculated density of states for zero and nonzero Hund's coupling.}
\end{figure}

We again start by considering only the dominant $dd\sigma$ hopping in the SK parameters;
the calculated tight-binding spectrum is shown in Fig.~\ref{fig:ek}(a). In the metallic pyrochlore
Cd$_2$Os$_2$O$_7$, the Os$^{5+}$ ion has three $d$ electrons corresponding to
a half-filled band. The Fermi level lies at $\epsilon = 0$ for this filling fraction
and the resultant Fermi `surface' consists of three lines and four points at the boundary
of the Brillouin zone. The three Fermi lines are diagonals of the square surface at the 
zone boundary, while the Fermi points are located at the high symmetry L-point $\mathbf k_{\rm L} 
= \left(\frac{1}{2},\frac{1}{2},\frac{1}{2}\right)$ [see Figs.~\ref{fig:ek}(a) and (b)].

Interestingly, this unusual Fermi surface can be nested by three wavevectors
$\mathbf Q_1 = (1,0,0)$, $\mathbf Q_2 = (0,1,0)$ and $\mathbf Q_3 = (0,0,1)$.
In particular, the Fermi lines are topologically equivalent to three `circles' each
of which can be completely nested by one of the $\mathbf Q$ vectors \cite{chern}. 
To determine the optimal ground state we minimize the energy among all the spin orderings 
for which the non-interacting system has a divergent susceptibility.  
In other words, we introduce a variational amplitude for the uniform ordering with $\mathbf Q_0 = 0$
and each wavevector ${\mathbf Q}_i$ that leads to perfect nesting of the Fermi surface. 
Restricted to this particular set of magnetic structures, 
our simulated-annealing minimization yields a noncoplanar spin order characterized by
a single wavevector $\mathbf Q_0 = 0$; the magnetic unit cell is the same as the 
crystal one. Spins on the four inequivalent sites
point toward the corners of a tetrahedron. The so-called all-in-all-out structure
shown in Fig.~\ref{fig:m-order1}(a) is an example of the noncoplanar `tetrahedral'
order. The corresponding band structure is shown in Fig.~\ref{fig:ek}(c) for  
Hund's coupling ${\tilde J}_H = 0.05t$. A charge gap opens at the original Fermi energy, as can 
also be seen in the calculated density of states [Fig.~\ref{fig:ek}(d)].

The noncoplanar spin structure obtained above can be a strong candidate for
the magnetic order below $T_{MI}$ in Cd$_2$Os$_2$O$_7$. This simple $\mathbf q=0$
order also preserves the cubic symmetry. Experimentally, the metal-insulator
transition was found to be accompanied by a slight change in unit-cell volume
of less than 0.05\%. More importantly, no change in crystal symmetry was observed below $T_{MI}$.
Although the exact magnetic structure is yet unclear, the $\mathbf q=0$ noncoplanar
order is consistent with a recent $\mu$SR measurement \cite{koda}. Interestingly,
upon further cooling, an incommensurate spin density wave discontinuously develops 
below $T\approx 150$ K \cite{koda}. This might indicate
the breakdown of mean-field approximation deep in the insulating
phase where strong electron correlations  play a predominant role.

In summary, we have studied the DE model with $t_{2g}$ electrons
on the pyrochlore lattice. By taking into account the orbital-dependent hopping,
we showed that magnetic properties of spinels close to the metal-insulator transition 
can be understood using the picture of cross-linking Kondo chains coupled by localized moments. 
Our theory provides simple and elegant explanations for the unusual spin superstructure observed 
in several vanadium spinels. We also proposed a novel noncoplanar `tetrahedral' order
for the magnetic insulating phase of the pyrochlore Cd$_2$Os$_2$O$_7$.

{\it Acknowledgement.} We thank Y.~Kato, I.~Martin, V.~Pardo, N.~Perkins, 
and F. Rivadulla for useful discussions. Work at LANL was carried out under 
the auspices of the U.S.\ DOE contract No.~DE-AC52-06NA25396 
through the LDRD program. G.W.C. is grateful to the hospitality of CNLS at 
LANL and the support of ICAM and NSF grant DMR-0844115.


\begin{thebibliography}{99}
\bibitem{note1}
In obtaining ${\bar H}$ we have neglected the Hund's terms that correspond 
to a hopping of singlet pairs between different orbitals of the same ion.
\end{thebibliography}
\end{document}